\documentclass[12pt,preprint]{aastex}
\usepackage{spr-astr-addons}
\usepackage{times}
\usepackage{lipsum,amsmath,multicol}
\usepackage{amsmath,amssymb}

\RequirePackage{color}

\begin{document}

\title{Compact stars with linear equation of state in  isotropic coordinates}
\slugcomment{}
\shorttitle{Short article title}
\shortauthors{Autors et al.}

\author{Sifiso A. Ngubelanga} 
\affil{Astrophysics and Cosmology Research Unit, School of Mathematics, Statistics and Computer Science, University of KwaZulu-Natal, Private Bag X54001, Durban 4000, South Africa
}

\author{Sunil D. Maharaj}
\affil{Astrophysics and Cosmology Research Unit, School of Mathematics, Statistics and Computer Science, University of KwaZulu-Natal, Private Bag X54001, Durban 4000, South Africa.}

\author{Subharthi Ray}
\affil{Astrophysics and Cosmology Research Unit, School of Mathematics, Statistics and Computer Science, University of KwaZulu-Natal, Private Bag X54001, Durban 4000, South Africa.}
\maketitle


\begin{abstract} 
We present a new class of spherically symmetric spacetimes for matter distributions with anisotropic pressures in the presence of an electric field. The equation of state for the matter distribution is linear. A class of new exact solutions is found to the Einstein-Maxwell system of equations with an isotropic form of the line element. We achieve this by specifying particular forms for one of the gravitational potentials and the electric field intensity. We regain the masses of the stars PSR J1614-2230, Vela X-1, PSR J1903+327, 4U 1820-30 and SAX J1808.4-3658 for particular parameter values. A detailed physical analysis for the star PSR J1614-2230 indicates that the model is well behaved. 


\end{abstract}

\section{Introduction}\label{sec:intro}

We consider solutions of the Einstein-Maxwell system of equations for charged static spherically symmetric interior distributions which match to the Reissner-Nordstrom exterior spacetime. Charged compact objects in relativistic astrophysics where the gravitational fields are strong are described by solutions of the coupled Einstein-Maxwell system of equations. The studies of \cite{Iva} and \cite{Sha} show that the presence of the electric field affects the values of surface redshifts, luminosities and maximum masses of compact objects. \cite{Mak} and \cite{Kom, Kom1} highlight the fact that the electromagnetic field has an important role in describing the gravitational behaviour of stars composed of quark matter. Models constructed in this way will be useful in describing the physical properties of compact relativistic objects, gravastars, neutron stars, etc, with different matter distributions. There have been several investigations in recent years on the Einstein-Maxwell system of equations for static charged spherically symmetric gravitational fields. Some recent comprehensive treatments are those of \cite{Kie}, \cite{Fat} and \cite{Mud}.

The Einstein-Maxwell field equations will have different forms depending on the coordinates utilized. In most  papers researchers have used canonical coordinates with neutral matter  and isotropic pressure distributions. For comprehensive lists of known solutions to the field equations, refer to \cite{Del}, \cite{Finch} and \cite{Stephani}. Many of these known solutions are not physically acceptable. For solutions to be physically reasonable it is necessary that the metric functions  and the matter variables are regular and well behaved in the interior of the star. Causality of the spacetime structure must be maintained, the energy conditions should be satisfied and physical quantities (for example the mass of a dense star) should correspond with observations of astronomical objects. For examples of recent papers  with charge and isotropic pressures the reader is referred to \cite{Fat} and \cite{Mud}. The general case of charged matter distributions with anisotropic pressures has generated much interest in several recent investigations.  Some recent treatments are those of \cite{Maf} and \cite{Mah}.  It is important to note that isotropic coordinates have not been used as often as canonical coordinates.  A recent example of charged matter with isotropic pressures in isotropic coordinates is the \cite{PantN} model.

Realistic matter distributions require an equation of state. A barotropic equation of state requires that the pressure be a function of the energy density. The form of the barotropic equation of state can be linear, quadratic, polytropic or some other dependence. Some classes of exact solutions to the Einstein-Maxwell system have been found by imposing an equation of state in canonical coordinates. Models of charged and anisotropic stars with a linear equation of state are those of \cite{Maf1}, \cite{Sun, Sun1}, \cite{Mah} and \cite{Escu}. Charged compact objects with a linear equation of state have been generated by {\cite{Maf2}; their models prove to be good approximations of the astronomical objects PSR J1614-2230, PSR J1903+327, Vela X-1, SMC X-1 and Cen X-3. \cite{Fer} found a class of charged anisotropic solutions with a quadratic equation of state.  \cite{Mah1} and \cite{Maf}, respectively, found new models with charge and anisotropic pressures by imposing quadratic and polytropic equations of state. An uncharged strange quark star model with the quadratic equation of state was presented by \cite{Mal1}. Other possibilities for barotropic equations of state are the van der Waals models, a recent exact solution was found by \cite{Thi}, and extensions leading to the so called generalised van der Waals models by \cite{Mal}.

\cite{Del} in their comprehensive treatment pointed out that only a few successful attempts have been made to obtain classes of exact static solutions of the Einstein field equations for neutral perfect fluid spheres. They also observe  that only nine of their solutions are regular and well behaved without considering the restriction on the red shift. Only two solutions with isotropic coordinates in their analysis were shown to be well behaved, and these are by \cite{Nar} and \cite{Gold}. It was later shown by \cite{Simon} that the \cite{PantSah} solution is also regular and well behaved. In recent past years, well behaved solutions in isotropic coordinates with charge have been found by \cite{PantN} and \cite{Prad}. As far as we can ascertain most analyses in isotropic coordinates restrict the pressures to be isotropic. In this paper we present a new class of charged exact solutions in isotropic coordinates with anisotropic pressures.

Many of the references mentioned above mainly use canonical coordinates unlike our treatment which utilizes isotropic coordinates. We believe that the use of isotropic coordinates may provide some new insights, and possibly lead to new classes of exact solutions. In this paper we follow the approach of using isotropic coordinates with anisotropic pressures in the presence of the electric field. The objective of this paper is to generate new classes of exact solutions to the Einstein-Maxwell system of equations, by imposing a linear barotropic equation of state, that model a charged anisotropic relativistic body in isotropic coordinates. In section 2, we present the Einstein-Maxwell field equations for charged anisotropic fluid spheres in static spherically symmetric spacetimes. The field equations are written in the terms of isotropic coordinates, and then transformed to new variables suggested by \cite{Kustaano}. This system of equations in transformed form is easier to integrate and analyze. New classes of exact solutions are presented in section 3. In section 4 we study the physical properties of the new classes of exact solutions and regain masses for particular observed objects. In section 5 we analyse the physical features for parameters associated with the star PSR~J1614-2230. Some concluding comments are made in section 6.

\section{The model}
We  model a dense general relativistic star with strong gravity. The metric of the interior spacetime in isotropic coordinates can be written as

\begin{equation}
\label{eq:g1}  d s^{2} = -A^{2}(r)dt^{2} + B^{2}(r)[dr^{2} + r^{2} (d \theta^{2} + \sin^{2} \theta d \phi^2)],
\end{equation}
in coordinates $(x^a)=(t,r,\theta, \phi)$ and where $A(r)$ and $B(r)$ are metric quantities representing the gravitational field. Relativistic astronomical objects such as compact stars in astrophysical scenarios are consistent with the metric  (\ref{eq:g1}). The energy momentum tensor

\begin{eqnarray}
\label{eq:g2} T_{ij}&=&\mbox{diag}\left(-\rho-\frac{1}{2}E^2, p_r-\frac{1}{2}E^2, p_t+\frac{1}{2}E^2, \right. \nonumber\\
&& \left. p_t +\frac{1}{2}E^2\right),
\end{eqnarray}
describes an anisotropic charged matter distribution. In (\ref{eq:g2}), $\rho$ is the energy density, $p_{r}$ is the radial pressure, $p_{t}$ is the tangential pressure and $E$ is the electric field intensity. These quantities are measured in terms of a timelike  unit four-velocity ${\bf u}$ where $u^i=\frac{1}{A}\delta^i_{0}$.

The Einstein-Maxwell field equations for the line element (\ref{eq:g1}) and matter distribution (\ref{eq:g2}) can be expressed as

\begin{eqnarray}
\label{eq:g3} 8\pi \rho + \frac{1}{2} E^{2} &=& -\frac{1}{B^{2}}\left[2 \frac{B''}{B}-\frac{B'}{B}\left(\frac{B'}{B}-\frac{4}{r}\right)\right],  \\  
\label{eq:g4}    8\pi p_{r} -\frac{1}{2} E^{2} &=& 2 \frac{A'}{A}\left(\frac{B'}{B^{3}}+ \frac{1}{r} \frac{1}{B^{2}}\right) \nonumber \\
																	& & + \frac{B'}{B^{3}}\left(\frac{B'}{B}+ \frac{2}{r}\right),    \\           
\label{eq:g5}   8\pi p_{t} + \frac{1}{2} E^{2}&=& \frac{1}{B^{2}}\left(\frac{A''}{A} + \frac{1}{r} \frac{A'}{A}\right) \nonumber \\
																	& & +\frac{1}{B^{2}}\left[\frac{B''}{B}-\frac{B'}{B}\left(\frac{B'}{B}-\frac{1}{r}\right)\right],    \\  
\label{eq:g6}  \sigma &=& \frac{1}{4\pi r^2} B^{-1} (r^2 E)',
\end{eqnarray}
in isotropic coordinates where a prime $(')$ denotes a derivative with respect to the radial coordinate $r$. The quantity $\sigma$ is the proper charge density. We utilize units where the speed of light $c=1$ and the Newton gravitational constant $ G= 1$. The system of equations (\ref{eq:g3})-(\ref{eq:g6}) governs the behaviour of the gravitational field for an anisotropic charged fluid in a static spherical field. From equations (\ref{eq:g4}) and (\ref{eq:g5}) we obtain the condition of pressure anisotropy which has the form 

\begin{eqnarray}
 \frac{A''}{A}+\frac{B''}{B} &=&  B^2 (8\pi\Delta + E^2) \nonumber \\
&&  + \left(\frac{A'}{A}+\frac{B'}{B}\right)\left(2\frac{B'}{B}+\frac{1}{r}\right), \label{eq:g7}
\end{eqnarray}

\noindent where the quantity $\Delta = p_{t}-p_{r}$ is the measure of anisotropy. For neutral matter with isotropic pressures $(E=0=\Delta)$, (\ref{eq:g7}) gives the condition of pressure isotropy in the form 

\begin{equation}
\label{eq:g8} \frac{A''}{A}+\frac{B''}{B} = \left(\frac{A'}{A}+\frac{B'}{B}\right)\left(2\frac{B'}{B}+\frac{1}{r}\right).
\end{equation}
A general algorithm producing new exact solutions to (\ref{eq:g8}), given a particular seed solution, 
was found by \cite{Ngu} after integrating a nonlinear Bernoulli equation.

The spacetime exterior to the charged matter distribution is given by 

\begin{eqnarray}
 \label{eq:g9}	     ds^2 &=&-\left(	1-\frac{2M}{R}+\frac{q^2}{R^2}\right)dt^2\nonumber \\
 && +\left(1-\frac{2M}{R}+\frac{q^2}{R^2}\right)^{-1}dR^2  \nonumber\\
			&&	+R^2(d\theta^2 + \sin^{2}\theta d \phi^2), 
\end{eqnarray}
in coordinates $(x^a) = (t, R, \theta, \phi)$. In (\ref{eq:g9}), $R$ is the radial coordinate of the exterior region, and $M$ and $q^2$ are the mass and charge of the ball, respectively, as determined by the external observer. The exterior spacetime is the Reissner-Nordstr\"om solution. By matching the first and second fundamental forms for the metrics (\ref{eq:g1}) and (\ref{eq:g9}) we obtain the junction conditions at the stellar surface. These conditions are given by 

\begin{eqnarray}
\label{eq:g10} 	A_{s} &=& \left(1-\frac{2M}{R}+\frac{q^2}{R^2}\right)^{\frac{1}{2}},\\	\nonumber\\
\label{eq:g11}	R_{s} &=& r_{s} B_{s},\\		\nonumber\\
\label{eq:g12}	\left(\frac{B'}{B}+\frac{1}{r}\right)_{s}r_{s}&=&  \left(1-\frac{2M}{R}+\frac{q^2}{R^2}\right)^{\frac{1}{2}},\\	\nonumber\\
\label{eq:g13}	r_{s}(A')_{s} &=& \left[\frac{M}{R}-\frac{q^2}{R^2}\right],
\end{eqnarray}
where the subscript ``s" denotes the surface of the star. The equations (\ref{eq:g10})-(\ref{eq:g13}) are the  boundary conditions in isotropic coordinates. Observe that equations  (\ref{eq:g11}) and  (\ref{eq:g13}) are equivalent to zero pressure of the interior solution on the boundary. 

The system of equations (\ref{eq:g3})-(\ref{eq:g6}) can be written in a different form by introducing a new variable $x$, and defining new functions $L$ and $G$, as follows

\begin{equation}
\label{eq:g14} x  \equiv  r^{2}, \hspace{0.5cm} L  \equiv  B^{-1},   \hspace{0.5cm}   G \equiv  LA. 
\end{equation}
The above transformation  (\ref{eq:g14}) has been used by other authors in spherically symmetric spacetimes as pointed out by \cite{Stephani} . Then the line element (\ref{eq:g1}) becomes 

\begin{equation}
\label{eq:g15}  ds^{2} = -\left(\frac{G}{L}\right)^{2}dt^{2} + \frac{1}{L^{2}}\left[ \frac{1}{4x}dx^{2} + x (d \theta^{2} + \sin^{2} \theta d \phi^2)\right].
\end{equation}
On applying the transformation (\ref{eq:g14}) to the field equations 
(\ref{eq:g3})-(\ref{eq:g6}) we generate the equivalent field equations

\begin{eqnarray}
\label{eq:g16} 8\pi \rho + \frac{1}{2} E^{2}   &=& 4[2xLL_{xx}-3(xL_{x}-L)L_{x}],\\                                                      
\label{eq:g17} 8\pi  p_{r}  -\frac{1}{2} E^{2}   &=& 4L(L-2xL_{x})\frac{G_{x}}{G}\nonumber \\
&& -4(2L-3xL_{x})L_{x},\\                                    
\label{eq:g18} 8\pi   p_{t}   + \frac{1}{2} E^{2}  &=& 4xL^{2}\frac{G_{xx}}{G}+4L(L-2xL_{x})\frac{G_{x}}{G} \nonumber \\
&& -4(2L-3xL_{x})L_{x}-8xLL_{xx},\\                                                                          
\label{eq:g19}  \sigma^2 &=& \frac{1}{4\pi^2 x} L^2  (E+xE_{x})^2,
\end{eqnarray}
where the subscript $``x"$ denotes differentiation with respect to the variable $x$. The condition of pressure anisotropy is now given by 

\begin{equation}
\label{eq:g20}		 \frac{G_{xx}}{G}-2\frac{L_{xx}}{L} = \frac{(8\pi \Delta + E^2)}{4xL^2}.   
\end{equation}   
In the new coordinates the mass function is 

\begin{equation}
\label{eq:g21}	m(x) = 2 \pi \int^x_{0}{\frac{1}{\sqrt{\omega}}\left[\omega \rho(\omega) +\frac{E^2}{8\pi}\right]d\omega},
\end{equation}
which represents the mass within a radius $x$ of the charged sphere.

A barotropic equation of state $p_{r}=p_{r}( \rho)$ should be satisfied by the matter distribution. We expect this to be the case  for a physically realistic relativistic star. For the simplest case we assume the linear equation of state 

\begin{equation}
\label{eq:g22} 	p_{r} = \alpha \rho - \beta,
\end{equation}
relating the radial pressure $p_{r}$ to the energy density $\rho$, and where $\alpha$ and $\beta$ are arbitrary constants.  Then with the linear equation of state (\ref{eq:g22}),  it is possible to express the system (\ref{eq:g16})-(\ref{eq:g19}) in the form 

\begin{eqnarray}
\label{eq:g23}	 8\pi	\rho   	 &=&	 4[2xLL_{xx}-3(xL_{x}-L)L_{x}]- \frac{1}{2} E^{2},\\                                                      
\label{eq:g24}	 	  p_{r}	 &=&	 \alpha \rho - \beta,\\                                    
\label{eq:g25}	 	  p_{t}	 &=& 	 p_{r}  +    \Delta,\\                                                              
\label{eq:g26}  8\pi 	\Delta 	 &=&	 4xL^2\frac{G_{xx}}{G}+\frac{8}{(1+\alpha)}L(L-2xL_{x})\frac{G_{x}}{G}\nonumber \\
&& -\frac{8(1+3\alpha)}{(1+\alpha)}xLL_{xx}\nonumber\\
					& &	+24xL^2_{x}-\frac{8(2+3\alpha)}{(1+\alpha)}LL_{x}+\frac{16 \pi \beta}{(1+\alpha)},\\ 	\nonumber\\
\label{eq:g27}	\frac{E^2}{2}&=&	\frac{8\alpha}{(1+\alpha)}xLL_{xx}-12xL^2_{x}+\frac{4(2+3\alpha)}{(1+\alpha)}LL_{x} \nonumber \\
&& +\frac{4}{(1+\alpha)}L(2xL_{x}-L)\frac{G_{x}}{G} -\frac{8\pi \beta}{(1+\alpha)},\\	
\label{eq:g28}	 \sigma^2	 &=& 	\frac{1}{4\pi^2 x} L^2  (E+xE_{x})^2.
\end{eqnarray}
In the system of equations (\ref{eq:g23})-(\ref{eq:g28}) we note that the equations are highly nonlinear in both potentials $L$ and $G$. This system contains the six matter variables ($\rho$, $p_{r}$, $p_{t}$, $\Delta$, $E$, $\sigma)$ and the two metric functions $(L, G)$. Clearly there are more unknown functions than independent field equations in the Einstein-Maxwell system. This suggests that we need to choose the form for two of the quantities in order to integrate and obtain some classes of exact solutions.

\section{Exact models}

We aim to generate exact models to the system of Einstein-Maxwell equations (\ref{eq:g23})-(\ref{eq:g28}) by choosing physically reasonable forms for the gravitational potential $L$ and the electric field intensity $E$. We make the specific choices

\begin{eqnarray}
\label{eq:g29}	L 	&=& a+bx,\\
\label{eq:g30}	E^2	&=& x(c+dx),
\end{eqnarray}
where $a, b, c$ and $d$ are real constants. We choose the gravitational potential $L$ to be a linear function and the electric intensity $E^2$ to be a quadratic function in the variable $x$. This ensures that the potential and the charge are finite at the centre and are regular in the interior. Then equation (\ref{eq:g27}) becomes the first order equation
 
\begin{eqnarray}
 \frac{G_{x}}{G} &=&  \frac{b(2+3\alpha)}{(a-bx)} \nonumber\\
 &&  -\left[\frac{16\pi \beta+(1+\alpha)(24b^2+c+dx)x}{8(a-bx)(a+bx)}\right], \label{eq:g31}
  \end{eqnarray}
in the potential $G$. On integrating (\ref{eq:g31}) we obtain 

\begin{equation}
\label{eq:g32} G(x) =   K(a-bx)^{\Psi}(a+bx)^{\Phi}e^{N(x)},
\end{equation}
where $K$ is the constant of integration. We have introduced 
the function $N(x)$ and  constants $\Psi$ and $\Phi$. These are given by 

\begin{eqnarray}
\label{eq:g33} N(x) &=& \frac{d(1+\alpha)x}{8b^2},\\	\nonumber\\
\label{eq:g34} \Psi &=& \frac{1}{16ab^3}\left\lbrace a(1+\alpha)(bc+ad) \right. \nonumber \\
&& \left. -8b^2[ ab(1+3\alpha)-2\pi \beta]\right\rbrace,\\ 
\label{eq:g35} \Phi &=& \frac{1}{16ab^3}\left\lbrace a(1+\alpha)[b(24b^2+c)-ad)]\right. \nonumber \\
&& \left. -16\pi b^2\beta\right\rbrace .
\end{eqnarray}
It should be noted in (\ref{eq:g33})-(\ref{eq:g35}) that $a\neq 0$ and $b\neq 0$. 

We can now find an exact solution to the Einstein-Maxwell system. The line element is given by 

\begin{eqnarray}
\label{eq:g36} ds^2 &=&   -K(a-br^2)^{2\Psi}(a+br^2)^{2(\Phi-1)}e^{2N(r)}dt^2\nonumber\\	   \nonumber\\
					& & 	+(a+br^2)^{-2}[dr^2+r^2(d\theta^2+\sin^2\theta d\phi^2)],
\end{eqnarray}
where $K$ is a constant of integration, the function $N(r)$ and the constants $\Psi$ and $\Phi$ are given by (\ref{eq:g33})-(\ref{eq:g35}). The quantities associated with the matter and the electric field have the form 

\begin{eqnarray}
\label{eq:g37}	8 \pi \rho    	 &=& 12ab-\frac{1}{2}r^2(c+dr^2),\\
\label{eq:g38}	p_{r}   	 &=&\alpha \rho-\beta,	\\
\label{eq:g39}	p_{t}   	 &=&	p_{r}+\Delta,\\		\nonumber\\
\label{eq:g40} 8 \pi 	\Delta  	&=& \frac{\Psi(a+br^2)r^2}{b(a-br^2)^2}\left[4b^3(\Psi -1)(a+br^2) \right. \nonumber \\
&& \left. -(a-br^2)\left\{ 8b^3\Phi+d(1+\alpha)(a+br^2)\right\}\right]\nonumber\\	
				& & +\frac{2}{(1+\alpha)}\left[8 \pi \beta+4b\Phi(a-br^2)\right. \nonumber \\
&& \left. -4b(a+br^2)\left\{2+3\alpha+\Psi\right\}\right]\nonumber\\
			& & +\frac{d(a+br^2)}{16b^4}\left[16b^2(a-br^2)+(1+\alpha)\right. \nonumber \\
&& \left. \left\{16b^3\Phi+d(1+\alpha)(a+br^2)\right\}r^2\right]\nonumber\\
			& &  +4b^2[6+\Phi(\Phi-1)] r^2,\\ \nonumber\\ 
\label{eq:g41}  \sigma^2         &=& \frac{\left(a+b r^2\right)^2\left(3c+4dr^2\right)^2}{16 \pi^2 (c+dr^2)},		\\	\nonumber\\
\label{eq:g42}		E^2    &=& r^2(c+dr^2).	
\end{eqnarray}
This exact solution is given in terms of elementary functions. The mass function has the form

\begin{equation}
\label{eq:g43}	m(r)     = \frac{1}{2} r^3\left[\frac{(12ab+c)}{3}-\frac{(c-2d)r^2}{10}-\frac{dr^4}{14}\right].
\end{equation}

\begin{table*}
\caption{Masses $m$, central density $\rho_{0}$, central radial pressure $p_{r0}$ and surface density $\rho_{s}$ of different stars corresponding to the parameters $b=0.5$, $c=0.01$, $d=0.01$ and $\alpha =0.931$.}
\begin{center}
\begin{tabular}{@{}lcccccc@{}}
\tableline
Star		&Observed mass $m$ & $a$ & $\rho_{0}$ & $p_{r0}$ & $\rho_{s}$   \\
\tableline
PSR  J1614-2230		& 1.97  & 1.96819 & 0.469871 & 0.000370433  & 0.469473 \\
Vela  X-1			& 1.77  & 1.76819 & 0.422124 & 0.000370433  & 0.421726 \\
PSR J1903+327		& 1.667 & 1.66519 & 0.397535 & 0.000370433  & 0.397137 \\
4U 1820-30			& 1.58  & 1.57819 & 0.376765 & 0.000370433  & 0.376367 \\
SAX J1808.4-3658	& 0.9   & 0.89819 & 0.214427 & 0.000370433  & 0.214029 \\
\tableline
\end{tabular}
\end{center}
\label{tab:5stars}
\end{table*}

\section{Physical features}\label{sec:newsoln}

The exact solution (\ref{eq:g37})-(\ref{eq:g42}) for the spacetime (\ref{eq:g36}) has a simple form and may be used to describe a charged anisotropic fluid sphere. In the stellar interior, the gravitational potentials, the matter variables and the electromagnetic variables are well behaved. It is important to observe that the electric field vanishes at the centre of the star. The matter density and proper charge density are finite at the stellar centre. At the stellar centre, $r=0$ we have 

\begin{eqnarray}
\label{eq:g44}		\rho_{0}	&=&	   \frac{3ab}{2\pi},\\	\nonumber\\
\label{eq:g45}		  p_{r0}	&=&	  \frac{3ab\alpha}{2\pi}-\beta ,
\end{eqnarray}
and the density and radial pressure are finite. The electromagnetic quantities have the finite values 

\begin{eqnarray}
\label{eq:g46}		\sigma^2_{0} &=& \frac{1}{c} \left(\frac{3ac}{4\pi}\right)^2,\\
\label{eq:g47}			E^2_{0}  &=& 0,
\end{eqnarray}
at the centre. The pressure anisotropy is given by 

\begin{eqnarray}
\label{eq:g48}		\Delta_{0} &=& \left(\frac{d}{8\pi}\right) \left(\frac{a}{b}\right)^2 \nonumber\\
&& +\frac{1}{\pi(1+\alpha)}[ab\Phi + 2\pi \beta -ab(2+3 \alpha + \Psi)],
\end{eqnarray}
at the centre. The gravitational potentials $A$ and $B$ are also finite at $r=0$ and therefore all relevant quantities are regular in the core of the star. For stability it is necessary that $\Delta$ vanishes at $r=0$; we show that this is possible in the next section for particular parameter values. The mass is finite and is affected by the presence of charge. 

The speed of sound is given by 

\begin{equation}
\label{eq:g49}		v = \left(\frac{dp_{r}}{d \rho}\right)^{0.5},
\end{equation}
which becomes $v=\alpha$. We must have $\alpha<1$ so that the velocity of sound is less than the velocity of light. In our analysis throughout we choose the value $\alpha=0.931$. This is the reasonable choice because compact stars are relativistic and $\alpha$ should be close to unity for these structures. Also the surface of a compact object should have zero radial pressure. This will ensure that the boundary conditions (\ref{eq:g10})-(\ref{eq:g13}) are satisfied and the metrics (\ref{eq:g1}) and (\ref{eq:g9}) match at the boundary. Since the equation of state is not polytropic we have a finite value of the density at the stellar surface. We obtain 

\begin{equation}
\label{eq:g50}	\rho_{s} = \frac{3ab}{2\pi} - \frac{1}{16\pi}(c+d),
\end{equation}
from equation (\ref{eq:g37}) in geometric units. In (\ref{eq:g50}) we have  fixed the radius of the star at $r=1$. Then from equation (\ref{eq:g38}) we see that the zero of the surface pressure constrains the constant $\beta$ by 

\begin{equation}
\label{eq:g51}		\beta = \alpha \rho_{s},
\end{equation} 
in terms of the surface density. 

We can now give certain numerical values to the various constants contained in our solutions. We have selected data from recent 
observations of five compact objects. These are PSR~J1614-2230 studied by \cite{demorest:2010}, Vela X-1 
analysed by \cite{rawls:2011}, PSR~J1903+3217 investigated by \cite{freire:2011}, 4U~1820-30 
studied by \cite{guver:2010} and SAX~J1808.4-3658 considered by \cite{elebert:2009}. We find that 
by varying the constant $a$ in equation (43) we regain the masses of these stars, keeping the other parameters fixed at $b=0.5$, $c=0.01$ and $d=0.01$. Notice that the stellar bodies are generally charge neutral, and hence we have imposed a smaller value of the parameters $c$ and $d$, responsible for bringing in charge into the system. Later on, in the detailed analysis of the star PRS~J1614-2230, we made a variation of these two parameters, to show how they change the electric field. In Table \ref{tab:5stars} we find the compact stars have observed masses varying from 0.9 $M_\odot$ to 1.97 $M_\odot$. In this table we have also included the corresponding values of the central density, central radial pressure and the surface density.  

The macroscopic stellar properties, represented by the mass and radius, are very important 
for the description of compact stars. It should be be noted that stellar masses are the derived parameters, obtained directly from observation. 
The stellar radius however, depend on the models chosen by researchers, and hence varies widely from model to model. 
In our work, in order to avoid the ambiguity of what the actual value of the radius should be, 
we have normalised the radial parameter, such that for $r=1$ we have the stellar surface.
This has the advantage of establishing  the surface density $\rho_s$ through equation ({\ref{eq:g50}).

Our model provides sufficient freedom to accommodate masses outside the range of objects
presented in Table \ref{tab:5stars}.
Another pulsar, PSR J0348+0432, with an observed mass of 2.01 $M_\odot$ has also been reported in 
the journal Science by \cite{antoniadis:2013}. In this treatment they measured the mass of the white dwarf partner 
to be 0.172 $M_\odot$ from spectral features, and used the radio timing data from the observation 
with the Green Bank Telescope, to derive such a high mass.

\section{The star PSR J1614-2230}
The parameter value $a=1.96819$ produces the mass 1.97 $M_\odot$ corresponding to the star PSR J1614-2230. We use this parameter to illustrate the variation of features for the matter, charge and gravity inside the star from the centre to the surface.

Table \ref{tab:anis} indicates the variation of density, radial pressure, tangential pressure and anisotropy in the stellar interior. The density and radial pressure are decreasing functions. The radial pressure vanishes at the boundary which is the requirement for a localised distribution of matter. The tangential pressure is well behaved. The anisotropy is finite and vanishes at the centre which is necessary for stability. Table \ref{tab:mass} gives the behaviour of the mass, electric field and charge density. The mass increases with increasing radius. The electric field and charge density are regular throughout the interior with $E=0$ at the centre. The effect of the charge is brought in through the constants $c$ and $d$. Tables \ref{tab:chrgd} and \ref{tab:chrgc} represent the total charge in the system so that $r=1$ is fixed at the boundary. It is evident that the effect of the parameter $d$ is more pronounced than that of the constant $c$ making the system substantially charged. Finally the gravitational potentials $A^2$ and $B^2$ are tabulated in Table \ref{tab:pot} for the set of parameters corresponding to PSR~J1614-2230 from the centre to the surface. The values obtained show that the potentials are finite and positive.

A graphical analysis also provides insight into the behaviour of the relevant quantities. We have plotted the density $\rho$ (Fig. \ref{fig:rho}), the radial pressure $p_{r}$ (Fig. \ref{fig:pr}), the tangential pressure $p_{t}$ (Fig. \ref{fig:pt}), the pressure anisotropy $\Delta$ (Fig. \ref{fig:delta}), the mass $m$ (Fig. \ref{fig:mass}), the electric field intensity (Fig. \ref{fig:elect}), the charge density (Fig. \ref{fig:chrg}) and the gravitational potentials (Fig. \ref{fig:potA} and Fig. \ref{fig:potB}). All the quantities are well behaved. The various figures have been plotted with the help of \cite{Wolfram}. \cite{Maf2} using a different approach also studied particular astronomical objects in general relativity with linear equation of state (\ref{eq:g22}). Our results in this paper are broadly consistent with their results.

\begin{table*}
\caption{Variation of energy density, radial pressure, tangential pressure and measure of anisotropy from the centre to the surface with parameters $a=1.96819$, $b=0.5$, $c=0.01$, $d=0.01$ and $\alpha =0.931$.}
\begin{center}
\begin{tabular}{@{}cccccccccc@{}}
\tableline
$r$ 		& $\rho$ & $p_{r}$ & $p_{t}$ & $\Delta$    \\
\tableline
 0		& 0.469871 & 0.000370433  & 0.000370433 & 0 \\
0.1		& 0.469869 & 0.000368562  & 0.0015641   & 0.00119554  \\ 
0.2 	& 0.469862 & 0.000362728  & 0.00521714  & 0.00485441 \\
0.3		& 0.469851 & 0.000352263  & 0.011554    & 0.0112017 \\
0.4		& 0.469834 & 0.000336057  & 0.0209783   & 0.0206422  \\
0.5 	& 0.469809 & 0.000312553  & 0.0341235   & 0.033811  \\
0.6		& 0.469773 & 0.000279751  & 0.0519384   & 0.0516586 \\
0.7		& 0.469726 & 0.000235207  & 0.0758281   & 0.0755929 \\
0.8		& 0.469662 & 0.00017603   & 0.107888    & 0.107712 \\
0.9		& 0.469579 & 0.0000988871 & 0.151303    & 0.151204 \\
1		& 0.469473 & 0            & 0.211052    & 0.211052 \\
\tableline
\end{tabular}
\end{center}
\label{tab:anis}
\end{table*}

\begin{table*}
\caption{Variation of mass, electric field intensity and charge density for charged bodies from centre to the surface with parameters $a=1.96819$, $b=0.5$, $c=0.01$ and $d=0.01$.}
\begin{center}
\begin{tabular}{@{}cccc@{}}
\tableline
$r$ 		 & $m$ & $E^2$ & $\sigma^2$   \\
\tableline
 0		& 0 		 & 0         & 0.00220779  \\
0.1		& 0.00196986 & 0.000101  & 0.00225602  \\ 
0.2 	& 0.015759   & 0.000416  & 0.00240346  \\
0.3		& 0.0531873  & 0.000981  & 0.00265829  \\
0.4		& 0.126075   & 0.001856  & 0.00303435  \\
0.5 	& 0.24625    & 0.0003125 & 0.0003557   \\
0.6		& 0.425518   & 0.004896  & 0.00423597  \\
0.7		& 0.675715   & 0.007301  & 0.00512147  \\
0.8		& 1.00866    & 0.010496  & 0.00624986  \\
0.9		& 1.43615    & 0.014661  & 0.00767248  \\
1		& 1.97       & 0.02      & 0.00945156  \\
\tableline
\end{tabular}
\end{center}
\label{tab:mass}
\end{table*}

\begin{table*}
\caption{Electric field intensity and charge density $(r=1)$ with parameters $r=1$, $a=1.96819$, $b=0.5$ and $c=0.01$.}
\begin{center}
\begin{tabular}{@{}ccc@{}}
\tableline
$d$ 		& $\sigma^2$ & $E^2$\\
\tableline
0		& 0.003472  & 0.01 \\
0.1		& 0.0648458 & 0.11 \\
0.2 	& 0.126554  & 0.21 \\
0.3		& 0.188272  & 0.31 \\
0.4		& 0.249994  & 0.41 \\
0.5 	& 0.311716  & 0.51 \\
0.6		& 0.373439  & 0.61 \\
0.7 	& 0.435163  & 0.71 \\
0.8 	& 0.496887  & 0.81 \\
0.9 	& 0.558611  & 0.91 \\
1      	& 0.620335  & 1.01 \\
\hline
\end{tabular}
\end{center}
\label{tab:chrgd}
\end{table*}

\begin{table*}
\caption{Electric field intensity and charge density $(r=1)$ with parameters $r=1$, $a=1.96819$, $b=0.5$ and $d=0.01$.}
\begin{center}
\begin{tabular}{@{}ccc@{}}
\hline
$c$ 		& $\sigma^2$ & $E^2$\\
\hline
0		& 0.00617245 & 0.01 \\
0.1		& 0.0405418  & 0.11 \\
0.2		& 0.0752451  & 0.21 \\
0.3		& 0.109959   & 0.31 \\
0.4		& 0.144676   & 0.41 \\
0.5		& 0.179394   & 0.51 \\
0.6		& 0.214113   & 0.61 \\
0.7		& 0.248832   & 0.71 \\
0.8		& 0.283552   & 0.81 \\
0.9		& 0.318271   & 0.91 \\
1		& 0.352991   & 1.01 \\
\hline
\end{tabular}
\end{center}
\label{tab:chrgc}
\end{table*}

\begin{table*}
\caption{Potentials $A^2$ and $B^2$ with varying radius with parameters $a=1.96819$, $b=0.5$, $c=0.01$, $d=0.01$ and $\alpha =0.931$.} 
\begin{center}
\begin{tabular}{@{}ccc@{}}
\tableline
$r$		& $A^2$ & $B^2$\\
\tableline
0		& 1.01316 & 0.258146 \\
0.1		& 1.04095 & 0.245514 \\
0.2		& 1.07452 & 0.233786 \\
0.3		& 1.11453 & 0.22288  \\
0.4		& 1.16189 & 0.212719 \\
0.5		& 1.21776 & 0.203237 \\
0.6		& 1.28374 & 0.194375 \\
0.7		& 1.36189 & 0.186081 \\
0.8		& 1.45505 & 0.178306 \\
0.9		& 1.56709 & 0.171009 \\
1		& 1.70349 & 0.164151 \\
\tableline
\end{tabular}
\end{center}
\label{tab:pot}
\end{table*}

\begin{figure}[h]
\begin{center}
\includegraphics[width=.45\textwidth]{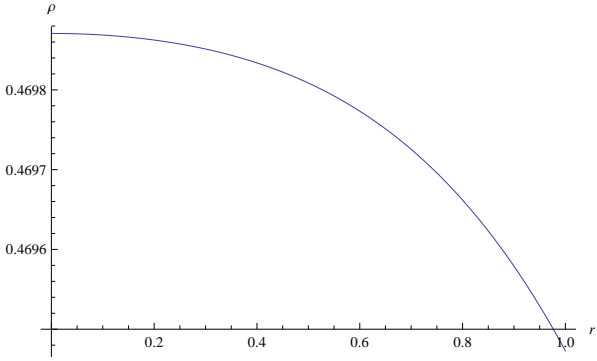}
\caption{Variation of density with the radius }
\label{fig:rho}
\end{center}
\end{figure}

\begin{figure}[h]
\begin{center}
\includegraphics[width=.45\textwidth]{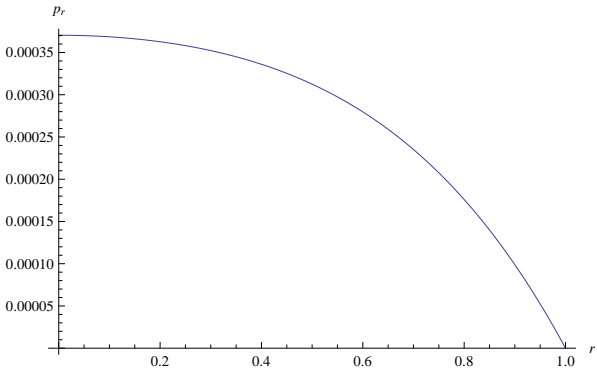}
\caption{Variation of radial pressure with the radius}
\label{fig:pr}
\end{center}
\end{figure}

\begin{figure}[h]
\begin{center}
\includegraphics[width=.45\textwidth]{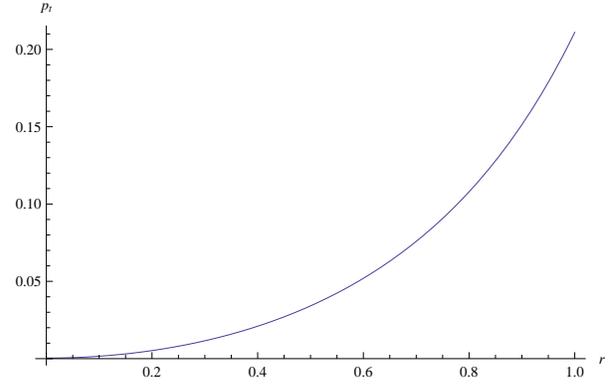}
\caption{Variation of tangential pressure with the radius}
\label{fig:pt}
\end{center}
\end{figure}

\begin{figure}[h]
\begin{center}
\includegraphics[width=.45\textwidth]{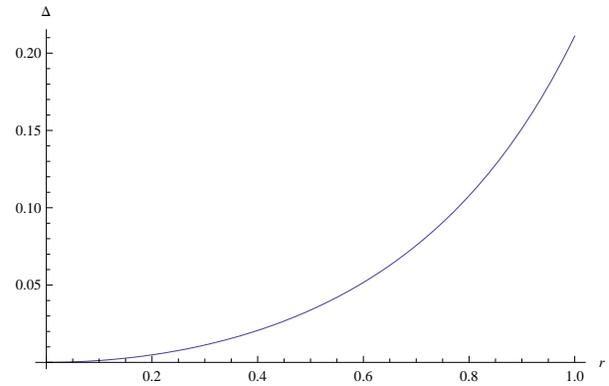}
\caption{Variation of the measure of anisotropy with the radius}
\label{fig:delta}
\end{center}
\end{figure}

\begin{figure}[h]
\begin{center}
\includegraphics[width=.45\textwidth]{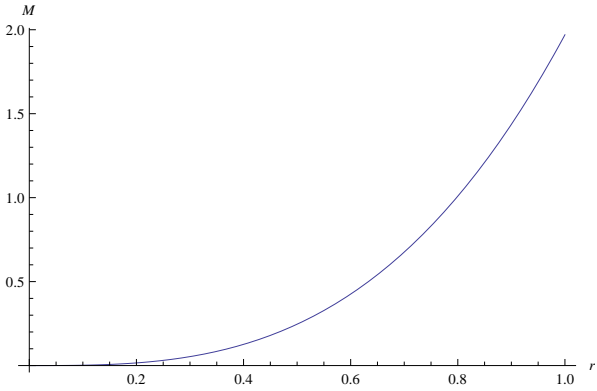}
\caption{Variation of mass with the radius}
\label{fig:mass}
\end{center}
\end{figure}

\begin{figure}[h]
\begin{center}
\includegraphics[width=.45\textwidth]{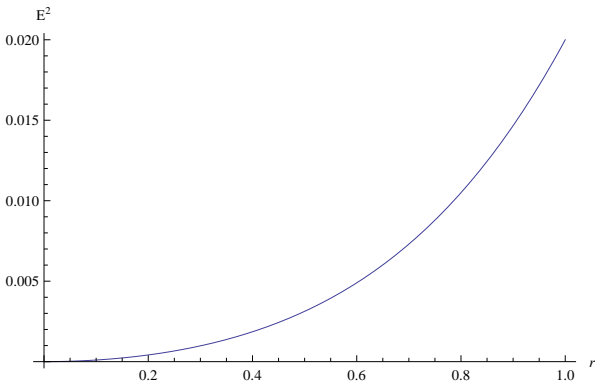}
\caption{Square of the Electric field intensity against the radius}
\label{fig:elect}
\end{center}
\end{figure}

\begin{figure}[h]
\begin{center}
\includegraphics[width=.45\textwidth]{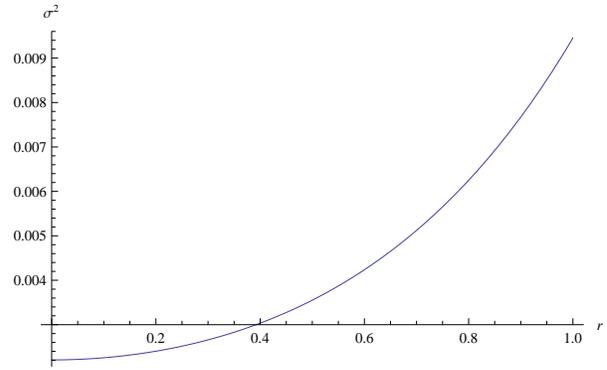}
\caption{Charge density against the radius}
\label{fig:chrg}
\end{center}
\end{figure}

\begin{figure}[h]
\begin{center}
\includegraphics[width=.45\textwidth]{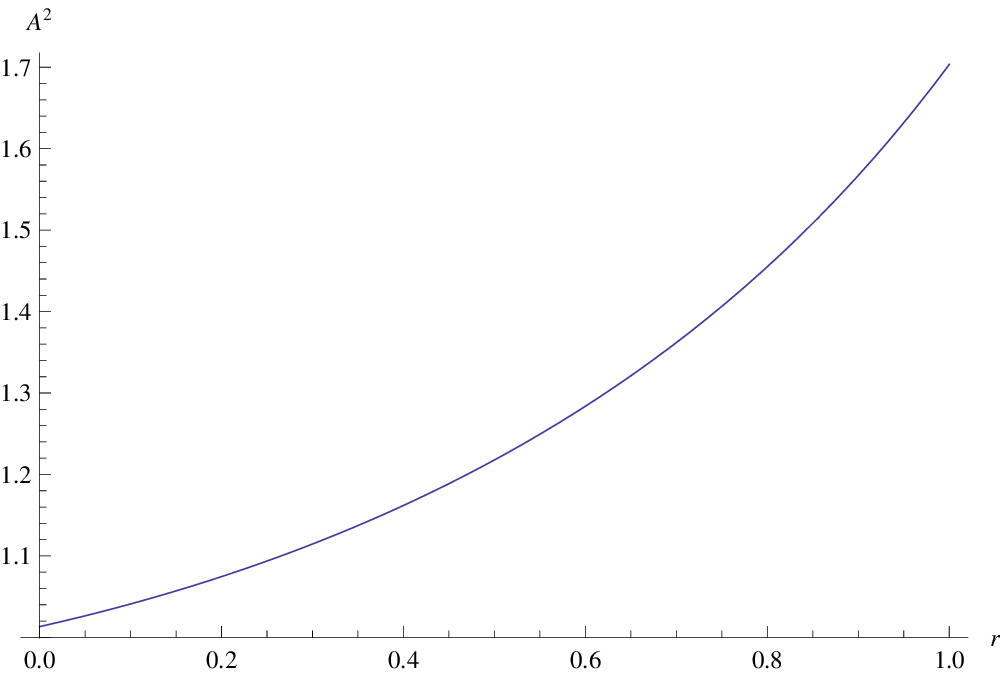}
\caption{Variation of the gravitational potential $A^2$ with the radius}
\label{fig:potA}
\end{center}
\end{figure}

\begin{figure}[h]
\begin{center}
\includegraphics[width=.45\textwidth]{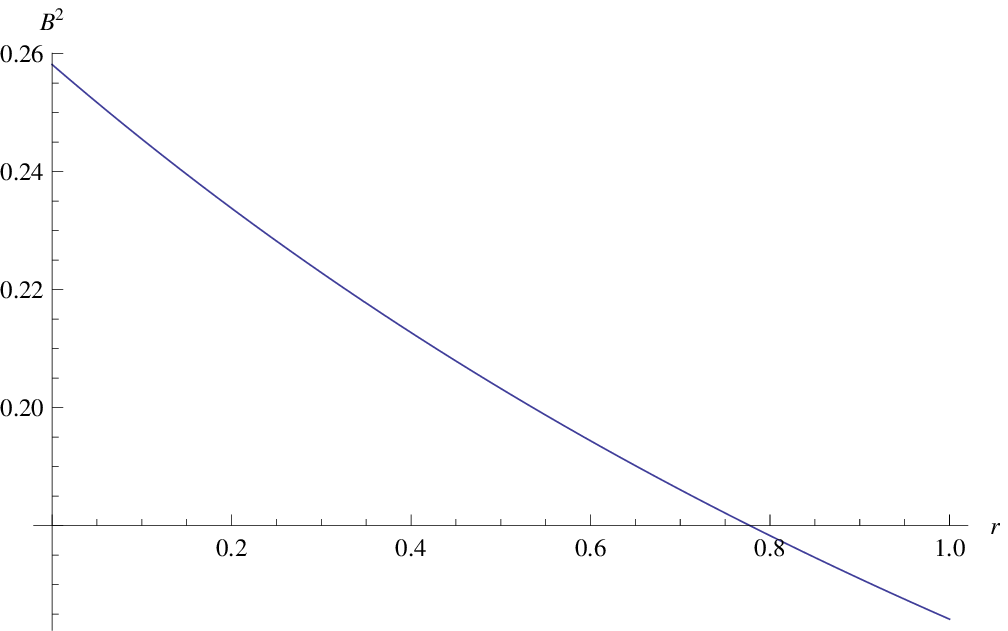}
\caption{Variation of the gravitational potential $B^2$ with the radius}
\label{fig:potB}
\end{center}
\end{figure}

\section{Discussion}

Our aim in this paper was to generate new exact models to the Einstein-Maxwell systems in isotropic coordinates for matter distributions with anisotropic pressures in the presence of charge. We imposed the barotropic equation of state which is linear and relates the radial pressure to the energy density. The exact solutions (\ref{eq:g37})-(\ref{eq:g42}) to the Einstein-Maxwell field equations generated are well behaved. We tabulated the matter and electromagnetic variables and showed that they are physically reasonable. By varying the parameter $a$ and choosing the fixed parameters $b=0.5$, $c=0.01$ and $d=0.01$ in Table \ref{tab:5stars}, we regain the masses of the stars PSR J1614-2230, Vela X-1, PSR J1903+3217, 4U 1820-30 and SAX J1808.4-3658. We used the star PSR J1614-2230 which has the mass 1.97 $M_\odot$   and fixed the parameter $a=1.96819$ to generate tables and graphical plots for gravitational potentials, matter variables and electromagnetic variables. We used Mathematica for graphical analysis with the particular parameter choices $a=1.96819$, $b=0.5$, $c=0.01$, $d=0.01$ and $\alpha=0.931$. The graphical analysis indicates that the model for PSR J1614-2230 is well behaved. The model has been generated by making the simple choices (\ref{eq:g29}) and (\ref{eq:g30}) for the potential and the charge respectively. Our approach may also be used to produce models which exhibit more general behaviour and thereby describe superdense stars with uncharged and charged matter. A different equation of state will also produce other qualitative features which are different from the linear case as shown in the treatment of \cite{Maf3}.

\end{document}